\documentclass{article}
\usepackage{authblk}
\usepackage{authblk}
\usepackage{float}
\usepackage{textcomp}
\usepackage{graphicx}
\usepackage{graphicx,subfig}

\usepackage{color}   
\usepackage{hyperref}
\hypersetup{
    colorlinks=true, 
    linktoc=all,     
    linkcolor=blue,  
}
\title{Studying the Supernova luminosity distance in Palatini formalism considering the role of causal structure constant}
\def\az{\mathcal{R}}
\def\palfR{$f(\az) $ }
\def\palfRR{$f(\az^{\mu\nu}\az_{\mu\nu}) $ }
\def\cST{$c_{ST} $ }
\def\dL{the luminosity distance }

\def\LIF{local inertial frame}

\author{\fontsize{11}{12}\selectfont Azam Izadi $^1$}
\author{Shadi Sajedi Shacker $^2$}
\affil{\fontsize{10}{11}\selectfont \itshape 1.Department of Physics, K. N. Toosi University of Technology, Tehran, Iran \\ \fontsize{10}{11}\selectfont \itshape 2.Hamburger Sternwarte, University of Hamburg, Hamburg, Germany}

\date{}
\usepackage{amsmath}
 \usepackage{multirow}
\usepackage{array}
\usepackage{amssymb}

\begin{document}
\maketitle

\begin{abstract}
The speed of light is a complicated synthesizer quantity with distinctive origins which lead to coincident values in the standard theory. 
Due to the fact that different aspects of speed of light do not coincide in the \LIF\ for different Palatini modified gravity theories, when deviating from general relativity, one should consider which aspect of speed of light should be taken into account meticulously and unambiguously. 
The aim of this study is mainly investigating the modification of the \textit{SN Ia} (Supernovae Type Ia) luminosity distance, for two \palfR and \palfRR extended theories in the \LIF\ in Palatini formalism considering different aspects of the speed of light. Besides the \LIF\ itself should be determined in the Palatini formalism as a frame in which the Einstein Equivalence Principle is valid. SN Ia luminosity distance should be modified considering the variation of the space-time causal structure constant for two well-known extended models in Palatini formalism.
 
\end{abstract}

\maketitle

\section{Introduction}

Nowadays the major challenge to the cosmological models is the late-time cosmic accelerating expansion of the universe. According to the cosmic observational evidences based on Type Ia supernovae (SN Ia) \cite{Riess, Perlmutter}, and standard rulers\cite{Shinji,palatini_gravity_theory}, our universe is experiencing an accelerating expansion phase.
Two general classes of models have been put forward in order to explain this expansion behavior. In the first class, it is attributed to the new gravitating component for the energy content of the universe, dark energy, with repulsive gravitational properties due to its negative pressure. 
The second class of models looks for an accelerating expansion via  modification of general relativity theory on cosmological scales or interpreting cosmological observations in another theoretical frame \cite{uzan}. A variety of approaches with richer space--time concepts are proposed to modify Einstein's theory of gravity. These can be put into different categories such as Scalar--Tensor theories, Tensor--Vector--Scalar theories (TeVeS), higher dimensions theories, Kaluza--Klein theory, and theories with a modified Lagrangian ($f(\az, \az_{\mu\nu}, \cdots)$) in three metric, Palatini, and metric--affine formulations. Among all these various options, the advantage of $f(\az)$ and $f(\az^{\mu\nu} \az_{\mu\nu})$ extended gravity theories is that no extra degree of freedom is suggested and the scalar curvature with well-understood physical origin causes the accelerating expansion. 
\\As it mentioned there are several formalisms in dealing with General Relativity such as \textit{metric formalism} and \textit{Palatini formalism}. 
In metric formalism Lagrangian is modified for extended theories and the connection is the  Christoffel symbol, but Birkhoff's theorem does not hold. 
In contrast, the metric and affine connection are regarded as two independent variables in Palatini formalism, the Birkhoff's theorem holds there and the energy--momentum tensor is taken independent of the connection  \cite{palatini_approach, palatini_approach_fR, fRgravity, palatini_gravity_theory}. Therefore, conservation of the energy--momentum tensor is still valid. 
\\It is worth noting that according to Ostrogradski theorem\cite{Ostro}, Lagrangians with higher derivatives  may suffer from instability problems. Based on Ostrogradski theorem, Hamiltonians containing more than first time derivative would have linear instabilities, on condition that the higher derivatives cannot be eliminated by partial integration. The result of Ostrogradski theorem is not relevant here since for the $f(\az)$ and $f(\az^{\mu\nu}\az_{\mu\nu})$ theories in Palatini formalism, the Lagrangian depends just on the first derivative of the fields.
\\The fact is that Palatini formalism gives a more generic geometric description of the space time.
In addition, there are some relations between some extended models such as Ricci scalar and Ricci squared theories in the Palatini version and non-perturbative approaches to quantum gravity, Loop Quantum Gravity, which motivate researchers to study Palatini extended theories.
Due to all these motivations, special attention has been paid to the Palatini formalism and its applications (e.g.\cite{tavakol_tsujikawa, Olmo_stellar, Binary}) in recent years.
In this paper we will consider two extended gravity theories, non-linear Ricci scalar and Ricci squared theories, in the framework of Palatini in which $\az_{\mu\nu}$ represents the Ricci tensor, which is made by a connection independent of the metric. The metric formalism can be obtained from Palatini one, assuming the simple form $f(\az_{\mu\nu})=\az_{\mu\nu}$ \cite{fRgravity}. 

As a matter of fact the cornerstone of general relativity theory is the \textit{equivalence principle}. The equivalence principle is true where one can find a local frame in which the gravitational effects are vanishing, which is called the \textit{local inertial frame}. 
Despite the fact that in the standard theory of gravity, a \LIF\ is considered a frame in which the metric is locally Minkowskian and therefore the Christoffel symbol is zero, this is a questionable case in the modified gravity theories in the Palatini formalism \cite{Izadi_1, OlmoPRL, open_problems_gr_phys}.
Since in such theories the connection is different from the Christoffel symbols, two different local frames can be defined in principle. One in which the connection is locally zero and the other in which the Christoffel symbols are locally vanishing. Therefore the main concern in the Palatini formalism is where the equivalence principle is true. These two frames are conformally related for Palatini $f(\az)$ gravity and thus the causal structure of both frames is the same, but this does not mean that they are completely equivalent from the physical point of view. 
For Palatini \palfRR gravity, these two frames are not conformally related and thus their causal structure is different. 
\\The point is that the action of a theory should be defined on some space–time with predefined properties. In the Palatini formalism the space–time structure is defined via both the metric (as the device for defining length) and the independent affine connection (as the device for parallel transporting). This means that in any experiment that the concept of parallelism is important, the effect of the independent connection can be seen and should be evaluated. This controversial subject was meticulously reconsidered in \cite{Izadi_1}. In the foregoing paper the \LIF\ in Palatini formalism, in which the equivalence principle holds, was chosen  wherever affine connection is locally zero. In \cite{Izadi_1} it is confessed that all the gravitational effects are removed from the equations in this introduced \LIF\ and then two samples of extended gravity theories were studied there. 
Since the Christoffel symbols are not zero, matter fields and the test particle trajectory are not the special relativistic ones, namely the particle's trajectory is not a straight line. For a general case, one can identify the extra terms appeared in the particle's trajectory with the force due to the variation of the speed of light \cite{Izadi_1}. 
\\On the other hand the speed of light, $c$, is a complicated synthesizer quantity which has different distinctive origins in Physics, which results in coincident values in standard theory. As a matter of fact $c$ is not the speed of ''light'' in all these theories.
Indeed, there is no persuasive reason to consider all different aspects of light speed such as \cST (the space-time causal structure constant), $c_{GW}$ (the gravitational wave velocity),  $c_{EM}$ (the electromagnetic wave velocity) and $c_{E}$ (the space-time--matter coupling constant appearing at the right hand side of Einstein's equations) the same in all formalisms \cite{Ellis} inasmuch as this fundamental constant enters many physical laws with different origins which are a prior unrelated. 
In any Deviation from GR or using a different formalism, one should consider precisely at first which aspect of speed of light is involved in the theoretical formulation and secondly whether all these facets still coincide and thirdly whether in an extended theory in different formalism, these facets remain constant or not. Furthermore the right technique of measuring spatial distances should be used \cite{Izadi_1, Izadi_2}.

On the whole, considering the advantages of Palatini extended theories and the fact that distinguishing different aspects of light speed may affect the analysis of cosmological data such as SN Ia, brings the idea of interpreting these data in a true \LIF\ . 
 
In what follows, the local inertial frame in Palatini formalism will be briefly introduced according to the result of \cite{Izadi_1}. We will see that for the foregoing extended models, working in palatini formalism leads to not coinciding different aspects of speed of light in the \LIF\.
\\The precise formulas of redshift, Hubble parameter and SN Ia luminosity distance will be rewritten by considering the right aspect of speed of light, \cST (coming from metric). Then we will show how the modifications, due to studying the extended Palatini theories in the right local inertial frame, affect the SN Ia luminosity distance in different \palfR and \palfRR models in section \ref{sec:redshift}. 
In section \ref{sec:Friedmannology}, the equations related to the SN Ia luminosity distance will be modified due to the variable feature of \cST in the \LIF\ in Palatini formalism. 
In section \ref{sec:proposed_model}, an appropriate model with a free parameter is proposed for modifying SN Ia relations to fit with observational data and its properties are studied as well.

\section{Local Inertial Frame in Palatini extended theories}

Two well-known extended gravity models, $f(\az)$ and $f(\az^{\mu\nu} \az_{\mu\nu})$ are under study in the present paper. 

\subsection{Palatini $f(\az)$ and $f(\az^{\mu\nu} \az_{\mu\nu})$ models}

In Palatini approach the Lagrangian density of the non-linear Ricci scalar gravity model is chosen to be an arbitrary function of the scalar curvature, $f(\az)$:
\begin{equation}
\mathcal{L} = \frac{1}{2\kappa} f(\az[g,\Gamma]) + \mathcal{L}_m
\label{action}
\end{equation}
\noindent in which $\mathcal{L}_m$ is the matter Lagrangian density and $\kappa=8\pi G/c_0^4$. It should be noted that the matter action does not depend on the connection in the Palatini formalism. 
The following equations of motion will be obtained by considering the metric and the connection as two independent variables:
\begin{equation}
f'(\az)\az_{\mu\nu}-\frac{1}{2}f(\az)g_{\mu\nu}=\kappa T_{\mu\nu}
\label{eq1}
\end{equation}
and
\begin{equation}\label{eq01}
D_{\alpha}\left (\sqrt{-g}f'g^{\mu\nu}\right )=0
\end{equation}
\noindent where $D_{\mu} X^{\nu}= \partial_{\mu}X^{\nu} + \Gamma^{\nu}_{\alpha\mu}X^{\alpha}$.
The above equation shows that by a conformal transformation, $h_{\mu\nu}=f'g_{\mu\nu}$, the connection is compatible with the metric $h_{\mu\nu}$ which is given by:
\begin{equation}\label{ggama}
\Gamma^{\alpha}_{\mu\nu}= {\alpha\brace \mu\nu}+ \gamma^{\alpha}_{\mu\nu}={\alpha\brace \mu\nu}+\frac{1}{2f'}\left [
2\delta^{\alpha}_{(\mu}\partial_{\nu)}f'-g_{\mu\nu}g^{\alpha\beta}.
\partial_{\beta} f'\right ]
\end{equation}

In Palatini $f (\az_{\mu\nu}\az^{\mu\nu})$ gravity, after ADM decomposition, Lagrangian density and the energy--momentum tensor are defined as below \cite{Izadi_1, Mota}.

\begin{equation}
 \mathcal{L} = \frac{1}{2\kappa}(\az + f(\az_{\mu\nu}\az^{\mu\nu})) +  \mathcal{L}_m
\end{equation}

\begin{equation}\label{Tmunu}
 T_{\mu\nu}=\rho u_\mu u_\nu + 2q _{(\mu} u_{\nu)} - p h_{\mu\nu} + \pi_{\mu\nu},
\end{equation}
\noindent in which $u^\mu = \frac{dx^\mu}{d\tau} $ is the 4-velocity normalized as $ u^\mu u_\mu=1$, $ h_{\mu\nu} = g_{\mu\nu} - u_\mu u_\nu$ which determines the orthogonal metric properties of observers moving with 4−velocity $ u^\mu$, 
$ \pi_{\mu\nu}= h^\alpha_\mu h^\beta_\nu T_{\alpha \beta}$ is the projected symmetric trace free anisotropic pressure, 
$ \rho = T_{\mu\nu} u^\mu u^\nu$ is the relativistic energy density relative to $ u^\mu$, 
$ q_\mu = h^\alpha_\mu u^\beta T_{\beta\alpha}$ is the relativistic momentum density, 
and $ p = \frac{-1}{3} h^{\mu\nu}T_{\mu\nu}$ is the isotropic pressure. 
In a similar way $\az_{\mu\nu}$ can be written as
\begin{equation} \label{Rmunu}
 \az_{\mu\nu} = \Delta u_\mu u_\nu + \Xi h_{\mu\nu} + 2 u_{(\mu}\gamma_{\nu)} + \varSigma_{\mu\nu}.
\end{equation}
Putting \eqref{Tmunu} and \eqref{Rmunu} in the modified Einstein equation, four equations are obtained for all the unknown coefficients.

The relation between the affine connection and the Christoffel symbol of physical metric is: 
\begin{equation}
\Gamma^{\alpha}_{\mu\nu}={\alpha\brace \mu\nu}+ \gamma^{\alpha}_{\mu\nu}
\label{18}
\end{equation}
\noindent where some of the above coefficients appears in $\gamma^{\alpha}_{\mu\nu}$ relation \cite{Izadi_1, Mota}. As it mentioned in the introduction section, there are different aspects of speed of light with different origins in physics. They will be considered in the next subsection. But it should be noted that in the \LIF\ \cite{Izadi_1} for this modified theory
\begin{equation}\label{cST}
 c_{ST} = \frac{c_0}{\sqrt{\omega}}.
\end{equation}
In order to find \cST in Palatini f($\az_{\mu\nu}\az^{\mu\nu}$) gravity one needs to know $\omega$ \cite{Izadi_1, Mota}:

\begin{equation} \omega=\frac{1+2F\Xi}{1+2F\Delta}. \end{equation}
F is the derivative of $f(\az^{\mu\nu}\az_{\mu\nu})$ with respect to $\az^{\mu\nu}\az_{\mu\nu}$ and $\Xi$ and $\Delta$ can be obtained from below equations:

\begin{equation}\label{delta}\Delta+2F\Delta^2 -\frac{1}{2}(\Delta+3\Xi+f)=\kappa\rho c_E^2 \end{equation} 
\begin{equation}\label{xi}\Xi+2F\Xi^2 -\frac{1}{2}(\Delta+3\Xi+f)=-\kappa p.\end{equation} 
\noindent Here we consider $c_E=c_0$ so $\kappa=\frac{8 \pi G}{c_0^4}$. 
Putting the right values for $\rho $ and $p$, $\omega$ can be obtained in different cosmological eras. 
If we take a simple model $f(\az^{\mu\nu} \az_{\mu\nu}) = a_1 \az^{\mu\nu} \az_{\mu\nu}$,

\begin{equation}
\label{omega}
 \frac{1}{\sqrt{\omega}} =
\begin{cases}
 \sqrt{\frac{1+3\sqrt{3+16 \kappa a_1 \rho_r^0 c_0^2  (1+z)^4}}{5-\sqrt{3+16 \kappa a_1  \rho_r^0 c_0^2 (1+z)^4}} }& \text{radiation dominated era} \\
\sqrt{\frac{1+\kappa a_1 \rho_m^0 c_0^2  (1+z)^3}{1-\kappa a_1 \rho_m^0 c_0^2  (1+z)^3}} & \text{matter dominated era.} 
\end{cases}
\end{equation}
And for the present de Sitter era $\omega\rightarrow 1$. 

 In addition, as the variation of the speed of light comes from the variation of energy content of the universe (energy momentum tensor),
 for local tests which can  be interpreted by Schwarzschild solution ($T_{\mu\nu}=0$), we have no variation.
 This means that our results are compatible with local varying speed of light tests such as the shift of the Mercury perihelion.
\subsection{Local inertail frame}

As we emphasized before the main concern in the Palatini formalism is where the equivalence principle is true or in other words in which local frame there is no gravity. In a general case the space-–time structure is defined by both the metric (as the device for defining length) and the independent affine connection (as the device for parallel transporting) in the Palatini formalism. This clearly means that in any experiment that the concept of parallelism is important, the effects of the independent connection should be examined. 
In principle one is able to define two local frames in Palatini approach. In the first one,  metric is locally Minkowskian, but the independent connection is not zero' namely $g_{\mu\nu}=\eta_{\mu\nu}$, ${\alpha\brace \mu\nu}=0$ and $\Gamma^{\alpha}_{\mu\nu}=\gamma^{\alpha}_{\mu\nu}$. There is also a second local frame in which the connection is zero and thus: $g_{\mu\nu}\neq \eta_{\mu\nu}$, $\Gamma^{\alpha}_{\mu\nu}=0$ and ${\alpha\brace \mu\nu}=-\gamma^{\alpha}_{\mu\nu}$. 
\\In order to understand the physical significance of the above mentioned frames, let's study the trajectory of a test particle in these two local frames. Hence we consider a dust with $T^{\mu\nu}=\rho u^{\mu} u^{\nu}$, where $\rho$ is the particle density and $u^{\mu}$ is the $4$--velocity. The energy--momentum conservation relation leads to the test particle trajectory. 
In the first local frame, the particle trajectory is a straight line 
\begin{equation}
\frac{d^2x^{\mu}}{ds^2}=0
\end{equation}
while in the second frame one has
\begin{equation}\label{trajectory}
\frac{d^2 x^\mu}{ds^2}-\gamma^\mu_{\alpha\beta} \frac{dx^\alpha}{ds} \frac{dx^\beta}{ds}=0,
\end{equation}
\noindent where the $\gamma^\mu_{\alpha\beta}$ is the difference of the affine connection and Christoffel symbol for both models. From the above equation we can see that the particle's trajectory is not a special relativistic one, straight line, in the second local frame.
\\Now the main question is that which one is the local inertial frame? 
\\It has to be noted that although in the first frame the particle moves on a straight line but this does not mean that equivalence principle is satisfied since gravity is present via the non--vanishing connection in other physical relations. So one expects to have changes in the geometrical concepts like geodesic deviation and Raychaudhuri's equation representing how a flux of geodesics expands. In this theory the existence of two different connection fields has new consequences. The geodesics are determined by the Christoffel symbols (this choice is motivated by energy--momentum conservation) but the equation that governs the evolution of the deviation vector involves the affine connection (motivated by the fact that the covariant derivative or parallel propagation along any arbitrary curve is defined by the affine connection). 
In the case that the Christoffel symbols are not zero, matter fields and the test particle trajectory are not the special relativistic ones. For a general case, one can identify the extra terms appeared in the particle's trajectory with the force due to the variation of the velocity of clock synchronization \cite{Izadi_1}. To see this, consider the case of Palatini $f(\az)$ gravity, in the second local frame (which is called the inertial frame). The observer at point 2 has to adjust its clock ahead with the amount $\sqrt{f'}d\ell/c_0$ (with $d\ell^2=f'|d\vec{x}|^2$). 
This shows that one should use the velocity $c_C=c_0/\sqrt{f'}$ for synchronization and as a result the test particle motion is given by:
\begin{equation}
\frac{d^2x^\mu}{d\tau^2}+\frac{dx^\mu}{d\tau}\frac{1}{c_C}\frac{dc_C}{d\tau}-\eta^{\mu\nu}c_0^2\frac{\partial_\nu c_C}{c_C}=0
\end{equation}
It is clear now that the extra terms are forces exerted on the particle because of the variation of the clock synchronization velocity. A similar discussion can be done for non-linear Ricci squared theories \cite{Izadi_1}. In standard theory the clock synchronization velocity coincides with the speed of light if we use the electromagnetic waves to do that.

In addition to the above discussions, we should notice the crucial point that the speed of light plays different key roles in physics and different aspects of the light speed may appear in various physical theories. It is a complicated synthesizer which has different distinctive origins, which result in coincident values in standard theory.
As it was shown in \cite{Izadi_1}, different aspects of light speed in $f(\az)$ and f($\az_{\mu\nu}\az^{\mu\nu}$) gravity became different from the amount $c_0$ ($\sim3\times10^8 m/{s^2}$), we normally consider for light speed in MKS units, in the \LIF\ . 
So whenever light speed appears in an equation, not only one should consider which kind of $c$ it is, but also the kind of extended gravity model must be known. 
For instance, \cST is not the same in Palatini $f(\az)$ and f($\az_{\mu\nu}\az^{\mu\nu}$) gravity in the \LIF . 
A summary of these attempts is shown in table \ref{speedt} \cite{Izadi_1}. 
According to the table \ref{speedt} if $\omega$ is known then \cST can be calculated in Palatini  $f(\az_{\mu\nu}\az^{\mu\nu})$ gravity. 
Also, it is important to imply that according to its behavior in these modified theories, \cST has been larger in the past than its present time value.  This would be a convincing motivator to investigate how a cosmological equation would be modified considering \cST has not always been $c_0$.

\begin{table} 
\caption{\label{speedt}Comparison of different speeds of light in the Palatini $f(\az)$ and f($\az_{\mu\nu}\az^{\mu\nu}$) theories.}
\begin{tabular}{|c| c| c|}
\hline                    
  &Palatini $f(\az)$ gravity & Palatini f($\az_{\mu\nu}\az^{\mu\nu}$) gravity  \\ [0.5ex] 
\hline                 
$c_{E}$ & $c_{0}$ & $c_{0}$   \\  
\hline 
$c_{GW}$ & $c_{0}$ & $c_{0}$  \\ 
\hline
$c_{EM}$ & varying & varying    \\ 
\hline
\cST & $c_{0}$ & varying$(\frac{1}{\sqrt{\omega}} c_{0})$ \\  [1ex]  
\hline 
\end{tabular}
\end{table}

At first this controversial subject was considered in \cite{OlmoPRL}. Olmo considered quantum applications of Palatini formalism for \palfR theories in Schwarzschild solution and shows that the equivalence principle is violated for microscopic applications \cite{OlmoPRL,Olmo_Hydrogen} in the local frame in which the metric is locally Minkowskian. He called this frame inertial.
He considered the microscopic applications.
However for macroscopic applications following the steps in \cite{OlmoPRL}, one can show as the metric outside the mass is Schwarzschild, the geodesic equations would be the usual ones in Schwarzschild case.
\\In macroscopic applications we meticulously reconsidered in \cite{Izadi_1} this important issue. In the foregoing paper the \LIF\ in Palatini formalism, in which the equivalence principle holds, was chosen  wherever affine connection is zero in such a way that all the gravitational effects are vanishing from the equations and then two samples of extended gravity theories were studied. 
Of course if one considers a Schwarzschild metric for the \palfR case, this would turn into a straight line. 
The main goal of this work is to examine the behavior of fundamental constants in the \LIF\ in Palatini formalism for two well-known extended gravity theories, $f(\az)$ and $f(\az^{\mu\nu})$ models, and considering the right aspect of speed of light in the relations which should predict the observational data.

\section{\label{sec:redshift} Changes in redshift}

Redshift is a cosmological parameter which most of the cosmological equations are expressed in term of. 
As the goal of this study is to rewrite SN Ia equations and interprete them in \LIF\ in Palatini formalism considering different facets of the speed of light,
here we tend to rewrite redshift and study the result.
For incoming null ray \cite{Coles}, 

\begin{equation}\label{fr}\int_{t}^{t_0}{\frac{cdt}{a(t)}}=-\int_{r}^{0} \frac{dr}{\sqrt{1-kr^2}}=f(r). \end{equation}
For a small time duration one can get

\begin{equation}\int_{t+\delta t}^{t_0+\delta t_0}{c\frac{dt}{a(t)}}= \int_{t}^{t_0}{c\frac{dt}{a(t)}}  \rightarrow c\frac{\delta t_0}{a_0(t)}=c\frac{\delta t}{a(t)},\end{equation}
and the conventional redshift is defined as

\begin{equation} \label{z} \frac{\nu_0}{\nu}=\frac{a(t)}{a_0(t)}=\frac{1}{1+z},\end{equation}
where $ a_0 (t)$ is the present scale factor. On the other hand, it should be noticed that in above equations light speed is considered constant c. But as the definition of redshift comes from line element $ds^2$ and according to \cite{Ellis}

\begin{equation} ds^2 = g_{\mu\nu} dx^{\mu}dx^{\nu}, \end{equation}
where $g_{00} = -c_{ST}^2$.
Equation \eqref{fr} must be rewritten as

\begin{equation}\label{newfr}\int_{t}^{t_0}{\frac{c_{ST} dt}{a(t)}}=-\int_{r}^{0} \frac{dr}{\sqrt{1-kr^2}}=f(r).\end{equation}
Due to the fact that in this equation speed of light is coming from space-time description of the line element of Special Relativity \cite{Ellis}, it is \cST. 
Therefore if one distinguishes between different kinds of light speed, redshift would take the form 

\begin{equation}\label{zprime}
1+z^\prime \equiv \frac{\nu}{\nu_0}=\frac{c_{ST}}{c_0}(1+z).
\end{equation}

It is obvious from the above equation that if in a formalism $c_{ST}=c_0$ then $z=z^{\prime}$. 
Otherwise, the redshift must be modified in each cosmological observation for the model which has been used. 
Specifically according to table \ref{speedt}, in both radiation and matter eras the above relation takes this form $(1+z^\prime)>(1+z)$ for $f(\az_{\mu\nu}\az^{\mu\nu})$  model.\\
In addition, if one wants to replace the frequency with its corresponding wavelength, $c_{EM}$ would be needed as it is the velocity of any electromagnetic wave \cite{Ellis}. 
In other words,

\begin{equation} \lambda=\frac{c_{EM}}{\nu} \nonumber \end{equation}
\begin{equation} 
\frac{\lambda_0}{\lambda}=\frac{c_{ST}}{c_{EM}} (1+z),
\end{equation}
in which $ c_{EM}$ is considered to approach $c_0$ in the present time.

Due to the modification of redshift, Hubble parameter will change as well. $H'$ has simple relation with $H$:
\begin{equation}\label{Hprime}
H'=H(1+\frac{1+z}{a_0} \frac{d}{dz}(Ln (c_{ST})).
\end{equation}
As a result redshift and Hubble parameter should be modified in extended gravity theories in Palatini formalism, when one takes into account different facets of the speed of light.

\section{\label{sec:Friedmannology}Extended Friedmannology}

\subsection{Type Ia Supernovae luminosity distance}
\subsubsection{Standard theory}

Before we proceed further, let us review some general points.
In standard cosmology, Luminosity distance $(d_L)$ is defined as \cite{review_on_DE}:

\begin{equation}
 d_L = f(\chi)\sqrt{\frac{L_s}{L_o}},
\end{equation}
in which $\sqrt{\frac{L_s}{L_o}} = (\frac{\Delta \nu_1}{\Delta \nu_0}) = (1+z)$ and $f(\chi) = \frac{c}{H_0\sqrt{\Omega^0_k}} sinh (\frac{\sqrt{\Omega^0_k}}{c} \int \frac{c dz}{E(z)})$.
Therefore,
\begin{equation}\label{dl}d_L=\frac{c}{H_0\sqrt{\Omega_k^0}}sinh(\sqrt{\Omega_k^0}\int_{0}^{z}{\frac{dz}{E(z)}})(1+z),\end{equation}
where density parameter is defined as$\Omega_k^0=\frac{-kc^2}{(a_0H_0)^2}$
and $k$ shows the curvature of the universe (open ($k=-1$), flat ($k=0$) or closed ($k=+1$)).
Also considering a universe without dark energy, the Hubble parameter can be shown by  

\begin{equation}
H^2=H_0^2  (\Omega_m^0(1+z)^3 + \Omega_r^0(1+z)^4 + \Omega_k^0 (1+z)^2).
\end{equation}

Here $H_0$ is the Hubble constant at the present time,
$\Omega_m^0$, $\Omega_r^0$, $\Omega_k^0$ are matter, radiation and curvature density parameters at the present time respectively.

Without considering dark energy, expanding the integral in \eqref{dl} results in
\begin{equation} d_L=\frac{c}{H_0}[z+\frac{1}{4}(1+\Omega_k^0)z^2+...].\end{equation}
Willing to consider the presence of dark energy, expanding the integral in \eqref{dl} results in \cite{Shinji}
\begin{equation}\label{usual_dL} d_L=\frac{c}{H_0}[z+\frac{1}{4}(1-3\omega_D\Omega_D^0+\Omega_k^0)z^2+...].\end{equation}

In order to explain the discrepancy of SN Ia luminosity distance, between theory and observations, this equation has to have a term like $-3\omega_D\Omega_D^0$.
This extra term comes from observations and it can be eliminated or less than this amount if a suitable theory is being proposed.\\
Furthermore from observational point of view, \dL is considered as \cite{Coles}

\begin{equation}\label{dL_observation}
 d_L = 10^{1+\mu/5} pc,
\end{equation}
in which $\mu= m-M $ is the distance modulus, m is the apparent magnitude and M is the absolute magnitude.
Also, as all equations are considered to be in m.k.s units here, $1 pc \simeq 3.08 \times 10^{16} m$.

\subsubsection{The effect of considering Palatini formalism on Supernovae type Ia data}
Here we are willing to put the right velocities according to their origins in the corresponding notions.
Rewriting the analogous relations, the first step is to specify which speeds are involved.
For Palatini formalism, $c$ gives its place to \cST in some equations; because when it refers to calculating lengths, one has to use the concept of space-time. 
In fact, if one is looking for an operational technique for measuring distances, using radar demands the modification in the line element, 
since the space-time velocity is varying in the \LIF\ in Palatini formalism \cite{Izadi_2}. 
What will be appeared in the line element is \cST, therefore the mentioned equations will get the forms 
$\sqrt{\frac{L_s}{L_o}} = (\frac{\Delta \nu_1}{\Delta \nu_0}) = (1+z')$ and 
$ f(\chi) = \frac{c_{ST}}{H'_0\sqrt{\Omega^0_k}} sinh (\frac{\sqrt{\Omega^0_k}}{c_{ST}} \int \frac{c_{ST} dz'}{E'(z')}) $.\\
As a result,

\begin{equation}\label{newdl} d_L=\frac{c_{ST}}{H_0^\prime\sqrt{\Omega_k^0}}sinh[\frac{\sqrt{\Omega_k^0}}{c_{ST}}\int_{0}^{z'}{c_{ST}\frac{d {z'}}{E'({z'})}}](1+z^\prime),\end{equation}

where $ E'(z')=\frac{H'}{H'_0}$ and $H'$ is introduced in \eqref{Hprime}.
Willing to consider low redshift region and using $sinh x$ expansion, one gets to

\begin{equation}d_L=(1+z^\prime)\int_{0}^{z'}{c_{ST}\frac{d {z'}}{H'}}.\end{equation}

\subsection{Palatini \texorpdfstring{\palfR}{Lg} gravity}

As it is shown in table \ref{speedt}, in the \LIF\ \cST is the same as $c_0$ in Palatini $f(\az)$ gravity. 
Hence rewriting equations in this model results in no change.

\subsection{Palatini \texorpdfstring{\palfRR}{Lg} gravity}

According to \eqref{zprime} and also table \ref{speedt}, considering $f(\az_{\mu\nu}\az^{\mu\nu})$ model in Palatini formalism
redshift would be:

\begin{eqnarray}
1+z^\prime &=& \frac{c_{ST}}{c_0}(1+z) \nonumber\\
&=&\frac{1}{\sqrt{\omega}}(1+z). 
\end{eqnarray}
According to \eqref{omega} it is obvious that \cST is greater in the past. 
Consequently,
\begin{equation}\label{zrelation}
\frac{1+z^\prime}{1+z}=\frac{1}{\sqrt{\omega}}=\begin{cases}
\sqrt{\frac{1+3\sqrt{3+16 \kappa a_1 \rho_r^0 c_0^2  (1+z)^4}}{5-\sqrt{3+16 \kappa a_1  \rho_r^0 c_0^2 (1+z)^4}} } & \text{radiation dominated era} \\
 \sqrt{\frac{1+\kappa a_1 \rho_m^0 c_0^2  (1+z)^3}{1-\kappa a_1 \rho_m^0 c_0^2  (1+z)^3}} & \text{matter dominated era.} 
\end{cases}
\end{equation}
This means that in Palatini f($\az_{\mu\nu}\az^{\mu\nu}$) gravity, interpreting redshift of a signal in \LIF\ leads in a greater value.
($z^\prime>z$)

Moreover, according to equation \eqref{newdl}, the luminosity distance would be rewritten in the form of:

\begin{equation} \label{palatinidl}d_L=\frac{c_0}{\sqrt{\omega}H'_0\sqrt{\Omega_k^0}}sinh[\sqrt{\omega}\sqrt{\Omega_k^0}\int_{0}^{z'}{\frac{1}{\sqrt{\omega}}\frac{dz'}{E'(z')}}](1+z').\end{equation}

It can be concluded here that rewriting the luminosity distance in this way, leads in some changes.
The exact difference can be specified by putting the right numerical values and a suitable model (see section \ref{sec:proposed_model}).
These results are indicative of larger luminosity distance for Palatini \palfRR compared to the Einstein de Sitter model and they emphasize on the power of Palatini formalism. 
The price of this achievement is varying some aspects of speed of light in the \LIF\ of extended gravity theories in Palatini formalism.

\section{\label{sec:proposed_model} Proposed model to have the most consistent modified luminosity distance for SN Ia}

Here the main goal is to fit a model in a way that fits the most with the concordance model ($\Omega^0_{\Lambda}=0.6825$ and $\Omega^0_{m}=0.3175$ \cite{Planck})
In what follows, a suitable model is being proposed in order to check the effect of varying \cST on SN Ia Luminosity distance. 

Considering flat universe in low redshift region, equation \eqref{palatinidl} becomes:

\begin{equation} \label{modeldl} 
d_L=(1+z')\int_{0}^{z'}c_{ST}\frac{dz'}{H'},
\end{equation}
in which \cST, $z'$ and $H'$ are defined in \eqref{cST}, \eqref{zprime} and \eqref{Hprime}.

If we choose a linear form of $f(\az_{\mu\nu}\az^{\mu\nu})=a_1 \az_{\mu\nu}\az^{\mu\nu}$ as \cite{Izadi_1}, $a_1$ whose dimension is length squared, can be used as the model's degree of freedom.
Choosing 
\begin{equation}\label{a1_model}
a_1 = \Lambda^\varepsilon l_p^{2(1+\varepsilon)}, 
\end{equation}
$\varepsilon$ can be set by the observational data.
In this equation, $\Lambda$ is the cosmological constant and $l_p$ is the Planck length.

\begin{figure}[ht!]
\includegraphics[scale=0.25]{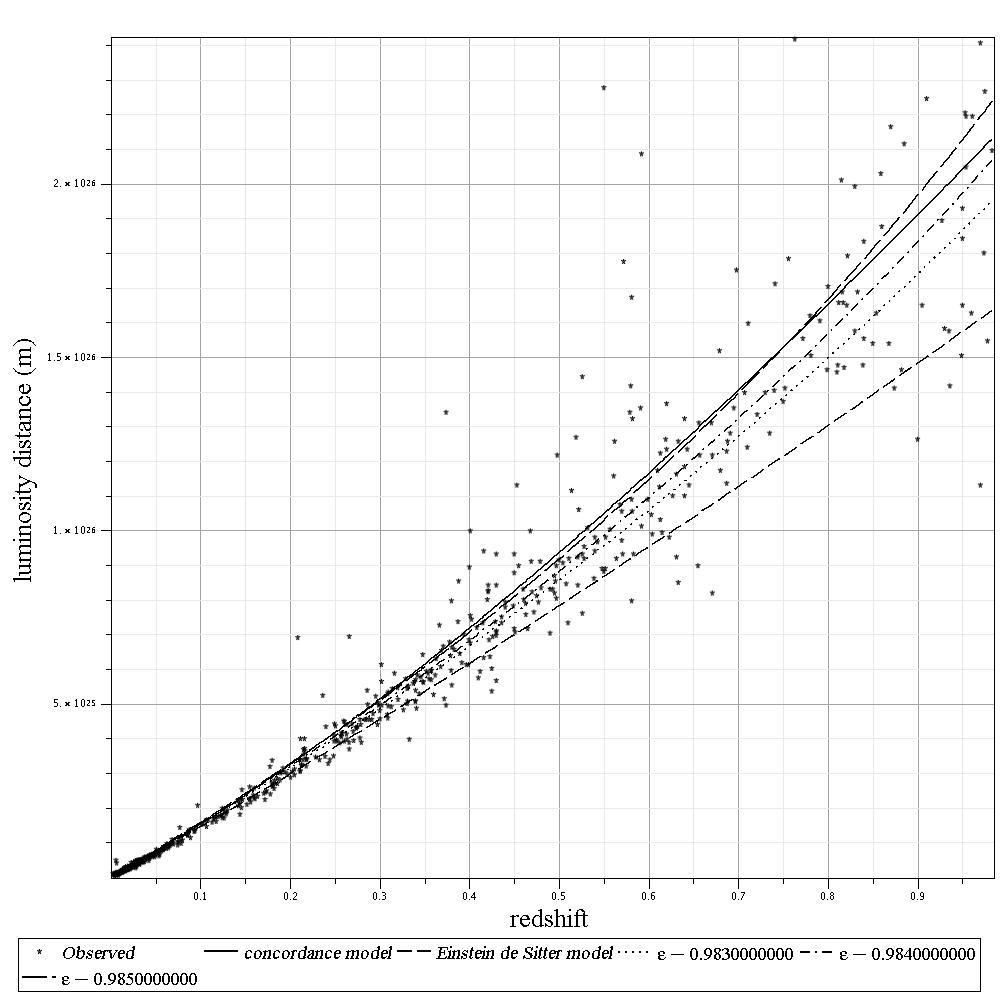}
\caption{\label{fig_dL}
The luminosity distance vs. redshift for
i) observations, using union 2 data for the distance modulus ($\mu$) \cite{website:union}.
ii) Einstein de- Sitter model, no dark energy is considered.
iii) concordance model, (considering dark energy $\Omega^0_{\Lambda}=0.6825$ \cite{Planck})
iv) Modified gravity in Palatini formalism considering the proposed model in equation \eqref{a1_model} and also distinguishing \cST from $c_0$ regarding the results in \cite{Izadi_1}.
As it is clear from the figure, $-0.987<\varepsilon<-0.98$ has the best fitting with the concordance model.
Compared to Einstein de-Sitter model, the modified gravity model in Palatini is more consistent with the concordance model.
In all models, the universe is considered to be flat.
}
\end{figure}

Studying more than 650 SN Ia (data from \cite{website:union}
) in $0<z<1$ region, 
gives a more or less similar value for $\varepsilon$, ($\varepsilon\sim-0.985$) in order to have a modified Luminosity distance which is more compatible with the observations compared to the Einstein de-Sitter model.
In fact, taking \palfRR model in Palatini formalism leads to larger luminosity distance with respect to metric formalism,
and \dL gets closer to the observational data.
In figure \ref{fig_dL} \dL is plotted for different values of $\varepsilon$
and is compared to models both with and without dark energy.

\begin{figure}[ht!]
\includegraphics[scale=0.21]{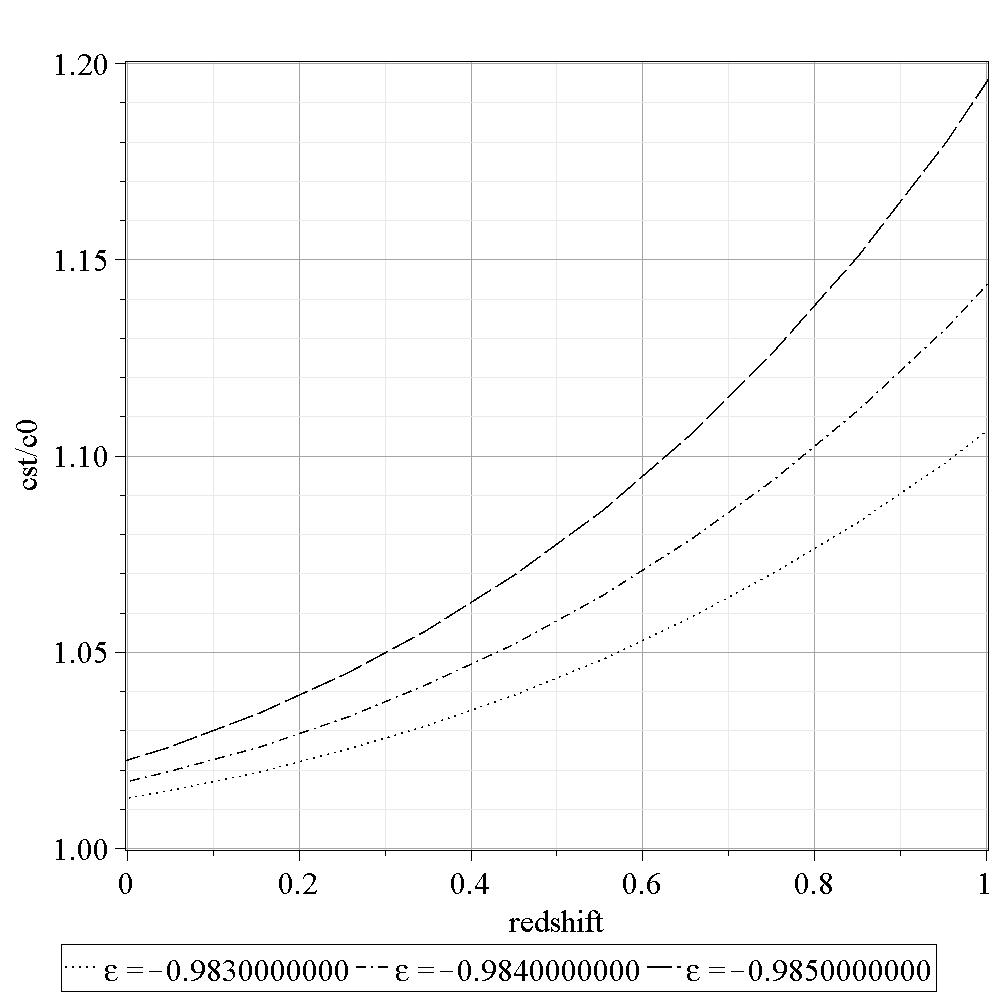}
\caption{\label{cST_for_985}Space time speed of light vs. redshift.
\cST in Palatini formalism for \palfRR model, increases with the redshift and for the present time it approaches to $c_0$
as for today's deSitter universe in $z=0$, $c_{ST}=c_0$.
}
\end{figure}

\begin{figure}[ht!]
\includegraphics[scale=0.21]{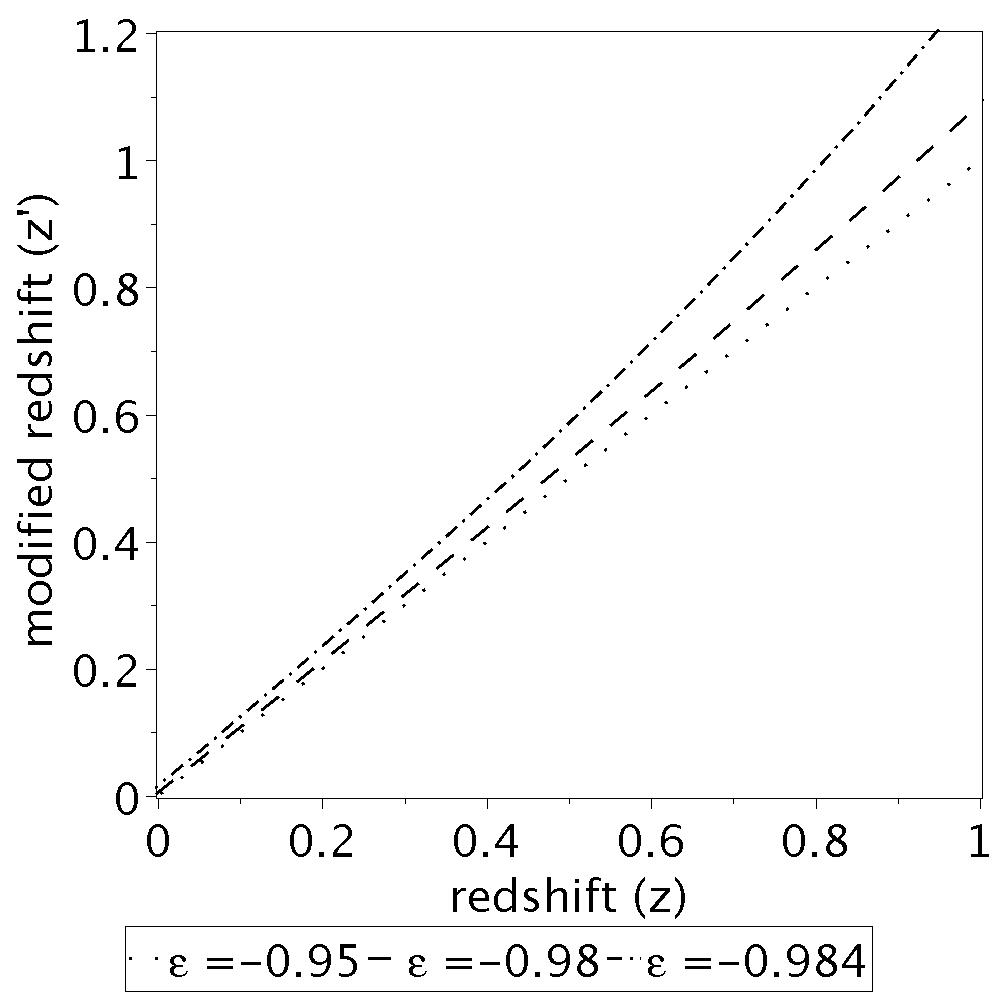}
\caption{\label{zprime_for_985}Modified redshift vs. redshift.
Considering modified gravity in Palatini formalism makes a shift in the value of the redshift.
Considering the right value of $\varepsilon$ regarding SN Ia luminosity distance, this graph is plotted for $\varepsilon=-0.985$.
This modification affects so many observational data.
Of course this value of $z$ is suitable for $0<z<1$ region and for higher redshift regions, more tests should be done (future work).}
\end{figure}

It is obvious from figure \ref{fig_dL} that the line with $\varepsilon\sim-0.985$ is the closest to the concordance model (in which $\Omega^0_{\Lambda}=0.6825$ and $\Omega^0_{m}=0.3175$ \cite{Planck}).
This leads to less need of dark energy for explaining the inconsistency between observations and theory for \dL.
Of course by testing this model on more cosmological equations, $\varepsilon$ can be fitted more precisely (future work).\\
Using the different values for $\varepsilon$, \cST and $z'$ are plotted in figures (\ref{cST_for_985}) and (\ref{zprime_for_985}).
Obviously for different values of $\varepsilon$, one gets different plots.

\section{Conclusion}
The speed of light plays a synthesizer role in physics.
Various aspects of speed of light appear in different physical theories with different origins \cite{Ellis}.
In standard theory the value of all these aspects such as \cST, $c_{EM}$, $c_E$ and $c_{GW}$ coincide; i.e. $c_{ST} = c_E = c_{GW} = c_{EM} = c_0$.
However considering modified gravity in Palatini formalism, all these aspects may not coincide at all.
In fact in the \LIF\ in which equivalence principle is satisfied, several forms of light speed such as \cST, $c_{EM}$, $c_E$ and $c_{GW}$ become different from each other considering modified gravity in Palatini formalism \cite{Izadi_1}. 
Therefore unlike the standard case, one cannot use $c_0$ as the speed of light in all theories. 
So in the determination of distance which needs both the measurement of time and a signal going from one point to another, it is important to distinguish between \cST and $c_{EM}$. 
All in all, for some modified gravity models in Palatini formalism, \cST may be varying in the \LIF\ and one should consider the effects of this change on the analysis of cosmological data such as the ones referring to dark energy.

For the redshift, rewriting equations leads to a new redshift which is indeed more than the usual one. 
This may considerably affect analysing observational data in which redshift plays an important role.
Hence using modifications above, redshift will have a larger value being interpreted in our choice of \LIF .

Finally, the SN Ia luminosity distance has been changed in Palatini formalism.
The modification is shown in equation \eqref{newdl}. 
Also a consistent model has been proposed. 
Within this model, there is a degree of freedom
and choosing the right value for it, can help in fitting the modified luminosity distance to cosmological observations.
It is shown in figure (\ref{fig_dL}) that if one chooses the right value for $\varepsilon$, the luminosity distance for the proposed model would fit better to the concordance model in $z \sim 1$
and there would be less need to dark energy.
One should also notice that for the local tests ($z=0$) using the Schwarzschild metric instead of FRW, \cST gets the value $c_0$.
In this model, the SN Ia modifications has been studied for the $0<z<1$ range.
Studying these effects in larger redshifts needs more details and considerations.
These considerations and also more tests such as the shift parameter of CMB and Baryonic Acoustic Oscillations are under study
so that $\varepsilon$ can be fitted with the observations the most.
\\The modifications are listed in table (\ref{alldata}).

\begin{table}[ht]
\caption{\label{alldata} Several observational data in two different formalisms: metric and Palatini. 
The standard Hubble constant $H$ is considered for Einstein de-Sitter model.
}
\begin{tabular}{|c|c|c|}\hline
&Metric formalism & Palatini gravity (\palfRR model) \\ 
\hline
redshift & $ z $ & $1+z^\prime=\frac{c_{ST}}{c_0}(1+z)$ \\  
\hline
$H$ & $ H_0 (1+z)^{3/2} $  & $H(1+\frac{d lnc_{ST}}{dz}(1+z))$ \\ 
\hline
$d_L$ & $ \frac{c}{H_0\sqrt{\Omega_k^0}} $  & $\frac{c_{ST}}{H'_0\sqrt{\Omega_k^0}}$ \\
&$\times sinh[\sqrt{\Omega_k^0}\int_{0}^{z'}{\frac{dz}{E(z)}}](1+z)$& $\times sinh[\frac{\sqrt{\Omega_k^0}}{c_{ST}}\int_{0}^{z'}{c_{ST}\frac{dz'}{E'(z')}}](1+z')$ \\ 
\hline
\end{tabular}
\end{table}

\textbf{Acknowledgments}: The authors are grateful to Robi Banerjee and Ali Shojai for their outstanding comments and also helpful discussions.
This work is supported by K. N. Toosi University of Technology.


\begin{thebibliography}{100}

\bibitem{Riess} S. Perlmutter, G. Aldering, G. Goldhaber, R. A. Knop, P. Nugent, P. G. Castro, S. Deustua, S. Fabbro, A. Goobar, D. E. Groom, I. M. Hook, A. G. Kim, M. Y. Kim, J. C. Lee, N. J. Nunes, R. Pain, C. R. Pennypacker, R. Quimby, C. Lidman, R. S. Ellis, M. Irwin, R. G. McMahon, P. Ruiz-Lapuente, N. Walton, B. Schaefer, B. J. Boyle, A. V. Filippenko, T. Matheson, A. S. Fruchter, N. Panagia, H. J. M. Newberg, W. J. Couch, and T. S. C. Project. Measurements of Ω and Λ from 42 High-Redshift Supernovae. Astrophys.J., 517:565–586, June 1999.

\bibitem{Perlmutter} A. G. Riess, A. V. Filippenko, P. Challis, A. Clocchiatti, A. Diercks, P. M. Garnavich, R. L. Gilliland, C. J. Hogan, S. Jha, R. P. Kirshner, B. Leibundgut, M. M. Phillips, D. Reiss, B. P. Schmidt, R. A. Schommer, R. C. Smith, J. Spyromilio, C. Stubbs, N. B. Suntzeff, and J. Tonry. Observational Evidence from Supernovae for an Accelerating Universe and a Cosmological Constant. Astron.J., 116:1009–1038, September 1998.

\bibitem{Shinji} L. Amendola and S. Tsujikawa. Dark Energy: Theory and Observations. 2010.

\bibitem{palatini_gravity_theory} M. Bastero-Gil, M. Borunda, and B. Janssen. The Palatini formalism for higher-curvature
gravity theories. In K. E. Kunze, M. Mars, and M. A. V ́azquez-Mozo, editors, American Institute of Physics Conference Series, volume 1122 of American Institute of Physics Conference Series, pages 189–192, May 2009.

\bibitem{uzan} J.-P. Uzan. Variation of the Constants of Nature in the Early and Late Universe. In
C. J. A. P. Martins, P. P. Avelino, M. S. Costa, K. Mack, M. F. Mota, and M. Parry, editors, Phi in the Sky: The Quest for Cosmological Scalar Fields, volume 736 of American Institute of Physics Conference Series, pages 3–20, November 2004.

\bibitem{palatini_approach} G. J. Olmo. Palatini Approach Beyond Einstein’s Gravity. ArXiv e-prints, December 2011.

\bibitem{palatini_approach_fR} G. J. Olmo. Palatini Approach to Modified Gravity:. f(R) Theories and Beyond. International Journal of Modern Physics D, 20:413–462, 2011. 

\bibitem{fRgravity} T. P. Sotiriou and V. Faraoni. f(R) theories of gravity. Reviews of Modern Physics, 82:451–497, January 2010. 

\bibitem{Ostro} M. Ostrogradski, Mem. Ac. St. Petersbourg VI 4 385, 1850

\bibitem{tavakol_tsujikawa} S. Fay, R. Tavakol, and S. Tsujikawa. f(R) gravity theories in Palatini formalism: Cosmological dynamics and observational constraints. Phys. Rev. D, 75(6):063509, March 2007.

\bibitem{Olmo_stellar} G. J. Olmo, H. Sanchis-Alepuz, and S. Tripathi. Stellar structure equations in extended Palatini gravity. Phys. Rev. D, 86(10):104039, November 2012.

\bibitem{Binary} K. Enqvist, H. J. Nyrhinen, and T. Koivisto. Binary systems in Palatini f(R) gravity. Phys. Rev. D, 88(10):104008, November 2013.

\bibitem{Izadi_1} A. Izadi and A. Shojai. The speed of light in extended gravity theories. Classical and Quantum Gravity, 26(19):195006, October 2009.

\bibitem{OlmoPRL} G. J. Olmo. Violation of the Equivalence Principle in Modified Theories of Gravity. Physical Review Letters, 98(6):061101, February 2007.

\bibitem{open_problems_gr_phys} S. Capozziello and G. Lambiase. Open problems in gravitational physics. ArXiv e-prints, September 2014.

\bibitem{Ellis} G. F. R. Ellis and J.-P. Uzan. c is the speed of light, isn’t it? American Journal of Physics, 73:240–247, March 2005.

\bibitem{Izadi_2} A. Izadi and A. Shojai. Measurement of the space-time interval in modified gravity theories in Palatini formalism. General Relativity and Gravitation, 45:229–241, January 2013.

\bibitem{Mota} B. Li, J. D. Barrow, and D. F. Mota. Cosmology of Ricci-tensor-squared gravity in the Palatini variational approach. Phys. Rev. D, 76(10):104047, November 2007.    

\bibitem{Olmo_Hydrogen} G. J. Olmo. Hydrogen atom in Palatini theories of gravity. Phys. Rev. D, 77(8):084021, April 2008.

\bibitem{Coles} P. Coles and F. Lucchin. Cosmology: The Origin and Evolution of Cosmic Structure, Second Edition. July 2002.

\bibitem{review_on_DE} S. Tsujikawa. Dark energy: investigation and modeling. ArXiv e-prints, April 2010.

\bibitem{Planck} Planck Collaboration, P. A. R. Ade, N. Aghanim, C. Armitage-Caplan, M. Arnaud,
M. Ashdown, F. Atrio-Barandela, J. Aumont, C. Baccigalupi, A. J. Banday, and et al. Planck 2013 results. XVI. Cosmological parameters. Astron.Astrophys, 571:A16, November 2014.

\bibitem{website:union} Supernova cosmology project. 


\end{thebibliography}
\end{document}